# Electronic quantum coherence induced by strong field molecular ionization

Jinming Chen[1,4,5], Jinping Yao[1,†], Haisu Zhang[2], Zhaoxiang Liu[1,4], Bo Xu[1,4], Wei Chu[1], Lingling Qiao[1], Zhenhua Wang[3], Julien Fatome[2], Olivier Faucher[2], Chengyin Wu[6,7], and Ya Cheng[1,3,7,*]

[1]*State Key Laboratory of High Field Laser Physics, Shanghai Institute of Optics and Fine Mechanics, Chinese Academy of Sciences, Shanghai 201800, China*

[2]*Laboratoire Interdisciplinaire Carnot de Bourgogne (ICB), UMR 6303 CNRS-Université, Bourgogne-Franche Comté, 9 Ave. A. Savary, 21078 Dijon Cedex, France*

[3]*State Key Laboratory of Precision Spectroscopy, East China Normal University, Shanghai 200062, China*

[4]*University of Chinese Academy of Sciences, Beijing 100049, China*

[5]*School of Physical Science and Technology, ShanghaiTech University, Shanghai 200031, China*

[6]*State Key Laboratory for Mesoscopic Physics, School of Physics, Peking University, Beijing 100871, China*

[7]*Collaborative Innovation Center of Extreme Optics, Shanxi University, Taiyuan, Shanxi 030006, China*

[†]*jinpingmrg@163.com*

[*]*ya.cheng@siom.ac.cn*

# Abstract

The existence of electronic coherence can fundamentally change the scenario of nonlinear interaction of light with quantum systems such as atoms and molecules, which, however, has escaped from observation in the investigations of strong field nonlinear optics in the past several decades. Here, we report on the generation of electronic quantum coherence by strong field ionization of nitrogen molecules in an intense 800 nm laser field. The coherence is experimentally revealed by observing a resonant four-wave mixing process in which the two pump pulses centered at 800 nm and 1580 nm wavelengths are temporally separated from each other. The experimental observation is further reproduced by calculating the nonlinear polarization response of $N_2^+$ ions using a three-level quantum model. Our result suggests that strong field ionization provides a unique approach to generating a fully coherent molecular wavepacket encapsulating the rotational, vibrational, and electronic states.

In a strong light field, the Coulomb potential barrier of an atom or a molecule can be significantly lowered, thereby allowing the electron to be released by tunneling through the distorted potential barrier with a probability extremely sensitive to the laser intensity [1]. The non-perturbative nonlinear tunnel ionization has enabled generation of high-order harmonic waves as well as attosecond electron and optical pulses [2-4], non-equilibrium plasma [5], and lasers in air [6-14]. In particular, the air lasing observed recently has aroused great interest and its underlying mechanism is still under debate [15-22]. Unlike the other strong field phenomena in which the resonance effect makes only trivial or minor contributions, the air lasing is dominated by resonant transition or resonantly enhanced nonlinear processes [19,20,23,24]. Therefore, the air lasing, or more specifically speaking, the lasing action induced by strong field ionization of nitrogen molecules, represents an area largely unexplored so far.

In this Letter, we report on generation of electronic quantum coherence in nitrogen molecular ions directly through strong field ionization. The coherence is revealed by observing generation of resonant four-wave mixing (FWM) beams in $N_2^+$ ions with the two driver pulses being temporally separated from each other. The resonant FWM beams from $N_2^+$ ions appears as laser-like radiations at the transition wavelengths between the ground $B^2\Sigma_u^+$ state and the excited $X^2\Sigma_g^+$ state. It is noteworthy that although the electronic coherence generated in various atoms and molecules has been intensively investigated which has triggered a series of intriguing nonlinear phenomena such as electromagnetically induced transparency [25] and lasing without inversion [26], little attention has been paid on its role when investigating the interaction of strong fields with atoms or molecules at non-resonant wavelengths. However, if one attempts to ionize nitrogen molecules with an 800 nm strong laser field, the wavelength of the laser field will be resonant with the transition between the ground $X^2\Sigma_g^+$ state and the excited $A^2\Pi_u$ state, indicating that the electron coherence can be established in $N_2^+$ ions during the strong field ionization. Such quantum coherence can be of great use because of its implications to generation of novel coherent light sources [8-14], nonlinear optical spectroscopy [27-29], filamentation-based remote sensing [30], and even slow light [31] and quantum information storage [32].

Our experimental setup is schematically shown in Fig. 1(a). The MIR laser pulses (1580 nm, 110 μJ, 60 fs) were generated by an optical parametric amplifier (OPA, HE-TOPAS, Light Conversion Ltd.), which was pumped by an 800 nm, 40 fs Ti:sapphire laser (Legend Elite-Duo, Coherent, Inc.). The residual pump laser from the OPA (800 nm, 1.8 mJ, 40 fs) serves as the NIR laser pulses as sketched in Fig. 1(a). The two beams were collinearly focused with an $f$=20 cm focal lens (i.e., L1) into a gas chamber filled with 4-mbar nitrogen gas. The peak intensities of the NIR and MIR pulses in the gas chamber were measured to be $1.2\times10^{14}$ W/cm$^2$ and $6.5\times10^{12}$ W/cm$^2$, respectively. The two beams were parallel with each other in their polarization unless otherwise specified. Their relative delay was controlled by a motorized translation stage. The exiting beams from the gas chamber were collimated using an $f$=15 cm focal lens (i.e., L2). Then both the MIR and NIR laser beams were removed using a broadband dielectric mirror followed by a piece of blue glass to enhance the signal to noise ratio. The generated nonlinear optical radiations were focused into an imaging spectrometer (Shamrock 500i, Andor) for spectral analyses. Figure 1(b) illustrates the energy level diagram of the multiphoton transition in $N_2^+$ ions, which will be further discussed later.

Figure 2(a) shows the measured spectrum in the range between 365 nm and 445 nm as a function of the time delay between the NIR and MIR laser pulses. In all the spectra measured with the contribution from both the NIR and MIR pulses, the weak background spectrum generated by either the single NIR or the single MIR pulse has been subtracted. In Fig. 2(a), a prominent peak is observed around 400 nm. Specifically, the peak is featured with a broad spectrum when the two pulses are temporally overlapped, which results from the FWM process of $\omega_{NIR} + 2\omega_{MIR}$ as has been intensively investigated in many gaseous media [33]. Quite surprisingly, when the NIR and MIR pulses are well separated in the time domain, a narrow-bandwidth radiation at 391.4 nm was generated which can last up to several picoseconds. The wavelength of 391.4 nm corresponds to the P-branch bandhead of rotational transitions between $B^2\Sigma_u^+(v''=0)$ and $X^2\Sigma_g^+(v=0)$ states of $N_2^+$ ions, while the radiations located on the blue side of 391.4 nm originate from R-branch rotational transitions between the two states [34]. We notice that the narrow-bandwidth radiation in Fig. 2(a) disappears when the MIR pulse arrives before the NIR pulse (i.e., with a negative delay).

Figure 2(b) quantitatively compares the spectra measured at three time delays of $\tau_0 = 0$

ps, $\tau_1$ = 0.3 ps, and $\tau_2$ = 1.3 ps. At the delay of $\tau_0$, a strong FWM beam is observed whose broad spectrum covers multiple transition lines of $N_2^+$ ions including 391.4 nm, 427.8 nm, etc. At the delay of $\tau_1$, the non-resonant FWM beam is suppressed due to the temporal separation of the two pulses, whereas a strong narrow-bandwidth radiation at the wavelengths of 391.4 nm is generated. The narrow-bandwidth radiation generated at the delay of 0.3 ps is determined to be nearly two orders of magnitude stronger than the broadband FWM beam at the wavelength of 391.4 nm. It should also be stressed that the 391.4 nm radiation is enhanced by 3~4 orders of magnitude as compared with that produced by the 800 nm NIR pulse alone. At the longer time delay of 1.3 ps (i.e., $\tau_2$), the intensity of 391.4 nm radiation has been significantly reduced which becomes comparable to the spectral intensity of the FWM beam generated at the zero time delay. In addition, as shown in Fig. 2(c), the $N_2^+$ radiation at 391.4 nm begins to grow after zero delay and reaches its maximum at the time delay of 0.3 ps. After the delay of 0.5 ps, the 391.4 nm radiation shows a slow decay with a lifetime of ~10 ps. This feature is completely missing in the conventional FWM process which requires the temporal overlap of all the participating pulses as evidenced by the non-resonant FWM spectrum recorded at zero time delay.

To reveal the origin of the narrowband $N_2^+$ laser-like radiation, we further conducted an experiment with a non-collinear scheme in which the NIR and MIR pulses intersect at a crossing angle of 30 mrad. Figure 3(a) illustrates the spatially-resolved FWM beam measured using the imaging spectrometer. According to the phase-matching diagrams shown in the box, the FWM beam generated via $\omega_{NIR} + 2\omega_{MIR}$ will propagate at an angle of 15.1 mrad, which nicely agrees with the measurement result. Moreover, Fig. 3(b) illustrates the spatially-resolved spectrum of the 391.4 nm radiation at the delay of 0.3 ps. The emission angles at 391.4 nm are close to that of the non-resonant FWM beam around 400 nm, indicating the fact that both the broad- and narrow-bandwidth radiations are generated with the same phase matching mechanism. .

Another decisive proof of the FWM-like feature of the narrow-bandwidth radiation at 391.4 nm can be obtained by examining its polarization property. An integrating sphere equipped with a fiber was used to collect the generated nonlinear radiations for eliminating the polarization sensitivity of our spectrometer, then the polarization of the beams was examined with a Glan-Taylor prism. In the experiment, we set the polarization directions of the NIR and MIR beams in the horizontal and vertical directions, respectively, and the delay was set at 1.3 ps. Figure 3(c) shows that the polarization of the 391.4 nm radiation follows that of the NIR pulses. It is well known that the nonlinear radiations generated via the third-order process $\omega_{NIR} + 2\omega_{MIR}$ should be polarized along the direction of the NIR laser electric field. In addition, such nonlinear processes typically require the temporal overlap of the participating waves [35]. Apparently, this is not the case for the measurement in Fig. 3(c).

Based on the experimental observations mentioned above, we discuss the physical mechanism underlying the measured time-frequency features for the broad-bandwidth

FWM beam and the narrow-bandwidth laser-like radiation. First, the ionization probability of $N_2$ molecules is estimated to be <1% with our experimental parameters, thus the FWM beam should mainly originate from the nonlinear responses of neutral $N_2$ molecules. Since the wavelengths of both the NIR and MIR pulses are far from the electronic resonance in neutral $N_2$ molecules, the third-order nonlinear polarization corresponding to the broadband FWM beam is only present within the temporal overlap of the two pulses, as observed in our experiment. Second, the NIR laser spectrum covers the one-photon resonance wavelength (i.e., 787.5 nm) between $X^2\Sigma_g^+(v=0)$ and $A^2\Pi_u(v'=2)$ states of $N_2^+$ ions, and the MIR laser spectrum covers the two-photon resonance wavelength (i.e., 1556.3 nm) between $A^2\Pi_u(v'=2)$ and $B^2\Sigma_u^+(v''=0)$ states. This suggests that the transition from $B^2\Sigma_u^+(v''=0)$ to $X^2\Sigma_g^+(v=0)$ states may be realized by an efficient resonant FWM process as conceptually illustrated in Fig. 1(b), giving rise to the strong coherent radiation at 391.4 nm. Additionally, the 391.4 nm radiation fulfills the same polarization and phase-matching conditions as the conventional FWM process. An interesting question is that how the NIR laser is connected with the MIR laser in this process as the two pulses are temporally separated from each other.

To gain insight into the experimental observations, we calculate the nonlinear polarization induced by the NIR and MIR laser fields using a simplified three-level model. In the model, three electronic states of $N_2^+$ ions, i.e., $X^2\Sigma_g^+(v=0)$, $A^2\Pi_u(v'=2)$, and $B^2\Sigma_u^+(v''=0)$, are taken into consideration, and the nuclear dynamics are ignored. The density matrix $\rho$ for the three-level system can be written as:

$$\rho = \begin{pmatrix} \rho_{11} & \rho_{12} & \rho_{13} \\ \rho_{21} & \rho_{22} & \rho_{23} \\ \rho_{31} & \rho_{32} & \rho_{33} \end{pmatrix} \qquad (1),$$

where $\rho_{11}$, $\rho_{22}$, and $\rho_{33}$ denote the population probabilities of X, A, and B states, and $\rho_{21}$, $\rho_{31}$, and $\rho_{32}$ represent the coherence between X and A, X and B, and A and B states. The evolution of the density matrix is calculated by solving the Liouville-von Neumann equation shown below [36,37]:

$$\frac{d\rho(t)}{dt} = -\frac{i}{\hbar}[H_I(t), \rho(t)] + \left(\frac{d\rho(t)}{dt}\right)_{Coll} \qquad (2).$$

In the Eq. 2, $H_I$ is the Hamiltonian in the interaction picture, which is written as [38]:

$$H_I(t) = E(t) \cdot \begin{pmatrix} 0 & \mu_1 e^{-i\omega_1 t} & \mu_2 e^{-i\omega_2 t} \\ \mu_1 e^{i\omega_1 t} & 0 & 0 \\ \mu_2 e^{i\omega_2 t} & 0 & 0 \end{pmatrix} \qquad (3),$$

where $E(t)$ is the combined electric field of the NIR and MIR pulses, $\omega_1$ ($\omega_2$) the angular frequency of A-X (B-X) transition, and $\mu_1 = \mu_{XA} \sin\theta$ ($\mu_2 = \mu_{XB} \cos\theta$) the dipole coupling strength between A-X (B-X) states. Here, $\mu_{XA}$ ($\mu_{XB}$) is the transition dipole moment of A-X (B-X), and the angle $\theta$ between the molecular axis and the driver laser field is set at 45° in the simulation [19,20]. Besides, the collisional dissipation of the three-level system is phenomenologically included in $\left(\frac{d\rho(t)}{dt}\right)_{Coll}$ as $\frac{d\rho_{ii}}{dt} = -\frac{\rho_{ii}}{T_1}$ and $\frac{d\rho_{ij}}{dt} = -\frac{\rho_{ij}}{T_2}$ (i,j=1,2,3, i≠j), with $T_1$ and $T_2$ being the depopulation and the decoherence times, respectively, which are chosen based on our experimental results. Physically, the dissipation of $N_2^+$ in the laser-generated plasma should be mainly caused by the frequent collisions with free electrons, which induce a fast population decay through electron-ion recombination in few tens of picoseconds and an even faster dephasing of electronic coherence in few picoseconds [39,40]. It should be noted that these collisional timescales are much longer than the durations of the NIR and MIR pulses in our experimental conditions. The induced polarization $P(t)$ in $N_2^+$ can be expressed as follow by summing over the microscopic dipoles:

$$P(t) = -N\left(\mu_1 \rho_{21} e^{-i\omega_1 t} + \mu_2 \rho_{31} e^{-i\omega_2 t} + \text{c.c.}\right) \qquad (4),$$

where $N$ is the number density of $N_2^+$ ions. Then the generated field at $\omega_2$ (391.4 nm) is obtained by the corresponding spectral component of the induced polarization $\tilde{P}(\omega_2) = \int_{-\infty}^{\infty} P(t) e^{i\omega_2 t} dt \propto \int_{-\infty}^{\infty} \rho_{31}(t) dt$. The analytical expressions in the theoretical analyses are shown in the supplementary materials [41].

The simulation result for the narrow-bandwidth radiation near 391.4 nm is shown in Fig. 4. In the simulation, the laser parameters are chosen to be the same as the experimental parameters, and the initial population of $N_2^+$ ions is assumed to be all in the ground X state. We can clearly see that the 391.4 nm radiation first undergoes a rapid increase until the time delay of 120 fs, then it decays with a timescale depending on the decoherence time $T_2$. These results are in qualitative consistence with the experimental observations shown in Fig. 2(c). The 391.4 nm radiation generated at the longer delays arises from the retarded FWM process enabled by the interaction of the resonant A-X polarization left by the NIR pulse with the ensuing MIR pulse. The dynamics of A-state population and A-X polarization induced by the NIR pulse are also shown in the inset of Fig. 4, with the presence of fast Rabi oscillations in both terms and the evidence of a strong A-X polarization lasting after the NIR pulse. More details can be found in the supplementary materials [41].

The key role of the electronic coherence is also evidenced by observing the generation of the narrow-bandwidth radiation at 427.8 nm wavelength as indicated by the white arrow in Fig. 2(a). Unlike the 391.4 nm radiation, the 427.8 nm radiation cannot be generated with the two driver lasers centered at 800 nm and 1580 nm, thus the electronic coherence cannot be encoded in the 427.8 nm radiation. In such a case, the polarization contributed by the coupling between $B^2\Sigma_u^+(v''=0)$ and $X^2\Sigma_g^+(v=1)$ states can only exist through the multiphoton process of $\omega_{NIR} + 2\omega_{MIR}$ with the temporally overlapped laser pulses. However, if the wavelength of MIR laser is tuned to 1870 nm, both the one-photon and two-photon resonances, as mentioned above, can be assured again, as illustrated in Fig. 5(c). Consequently, the transition from $B^2\Sigma_u^+(v''=0)$ to $X^2\Sigma_g^+(v=1)$ states can now be realized by the resonant FWM process to generate the strong $N_2^+$ laser-like radiation at 427.8 nm. Likewise, the resonant interaction of the NIR laser with the $N_2^+$ ions generates a polarization between $X^2\Sigma_g^+(v=1)$ and $A^2\Pi_u(v'=3)$ states with a relaxation time much longer than the duration of the NIR pulse, enabling the generation of the 427.8 nm radiation even when the 800 nm and 1870 nm pulses are temporally separated by several picoseconds, as shown in Fig. 5(a) and (b). Specifically, the discrete peaks on the blue side of the 427.8 nm radiation originate from the R-branch rotational transitions between $B^2\Sigma_u^+(v''=0)$ and $X^2\Sigma_g^+(v=1)$ states. The oscillations in the R-branch radiations with the increasing time delay as observed in Fig. 5(a) reveal the quantum coherence of rotational wavepackets of $N_2^+$ ions [42].

To conclude, we have revealed the electronic quantum coherence in $N_2^+$ ions generated with the interaction of a strong 800 nm laser field with nitrogen molecules. Previous investigations also show that strong field photoionization of nitrogen molecules produces highly coherent rotational and vibrational wavepackets in $N_2^+$ ions [10,23,42]. The fully coherent molecular ion system provides opportunities for exploring new effects and applications in nonlinear and quantum optics.

**Acknowledgements**

This work is supported by National Basic Research Program of China (2014CB921303), National Key Research and Development Program (2018YFB0504400), National Natural Science Foundation of China (11734009, 61575211, 11625414 and 61590934), Strategic Priority Research Program of CAS (XDB16000000), Key Research Program of Frontier Sciences, CAS (QYZDJ-SSW-SLH010), Project of Shanghai Committee of Science and Technology (17JC1400400), Shanghai Rising-Star Program (17QA1404600), Program of Youth Innovation Promotion Association of CAS, the CNRS, the European Research Council under Grant Agreement 306633, ERC PETAL, the European Union (FEDER program), the Labex ACTION program (No. ANR-11-LABX-01-01). H. Z and J. F would like to acknowledge the support from the Conseil

Phys. Rev. A **89**, 042508 (2014).

**Captions of figures:**

Fig. 1 (Color online) (a) Schematic diagram of the experimental setup. (b) Energy diagram of the resonant FWM in $N_2^+$ ions for generating the 391.4 nm radiation.

Fig. 2 (Color online) (a) Spectrum measured as a function of the time delay between the NIR (800 nm) and MIR (1580 nm) laser pulses (logarithmic color scale). (b) Spectra captured at $\tau_0$ = 0 ps, $\tau_1$ = 0.3 ps and $\tau_2$ = 1.3 ps. (c) The evolution of 391.4 nm and 427.8 nm laser-like radiations and the non-resonant FWM beam near 400 nm with the increasing time delay.

Fig. 3 (Color online) (a) Spatially-resolved FWM spectrum measured at zero time delay. Phase-matching diagram of the FWM process is shown in the box. (b) Spatially-resolved spectrum of the 391.4 nm radiation measured at the delay of 0.3 ps. (c) Polarization properties of NIR pulses, MIR pulses, and 391.4 nm radiation characterized with a Glan-Taylor (G-T) prism.

Fig. 4 (Color online) Simulated 391.4 nm radiation as a function of the time delay between the 800 nm and the 1580 nm laser pulses. Inset: The evolution of the A-state population $\rho_{22}$ and corresponding A-X polarization $|\rho_{21}|$ induced by the NIR laser. For comparison, the envelope of the NIR laser field is indicated by black dot lines.

Fig. 5 (Color online) (a) Spectra measured as a function of the time delay between the 800 nm and 1870 nm laser pulses (logarithmic color scale). (b) The evolution of 427.8 nm radiation and the non-resonant FWM bream near 430 nm with the time delay. (c) Energy diagram of the resonant FWM for generating the 427.8 nm radiation.

Fig. 1

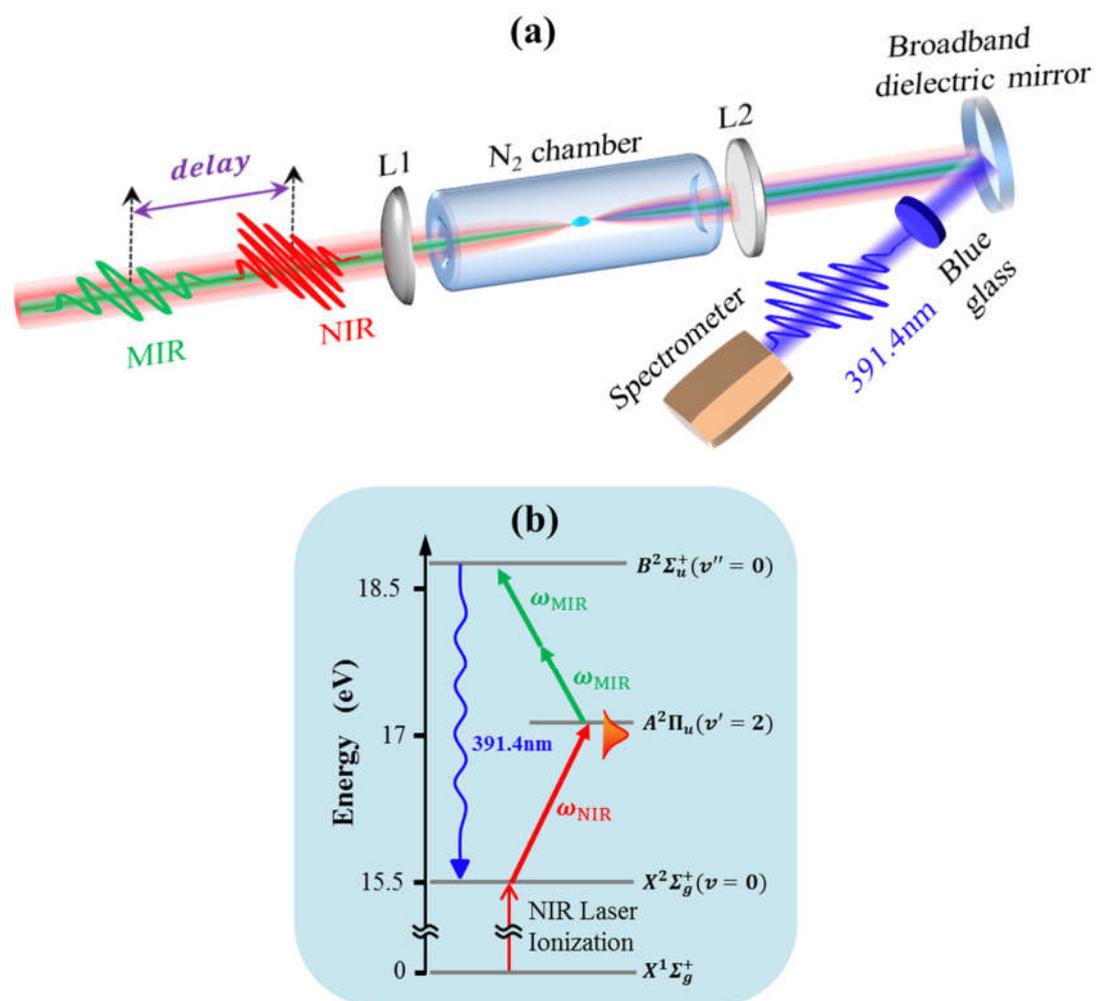

Fig. 2

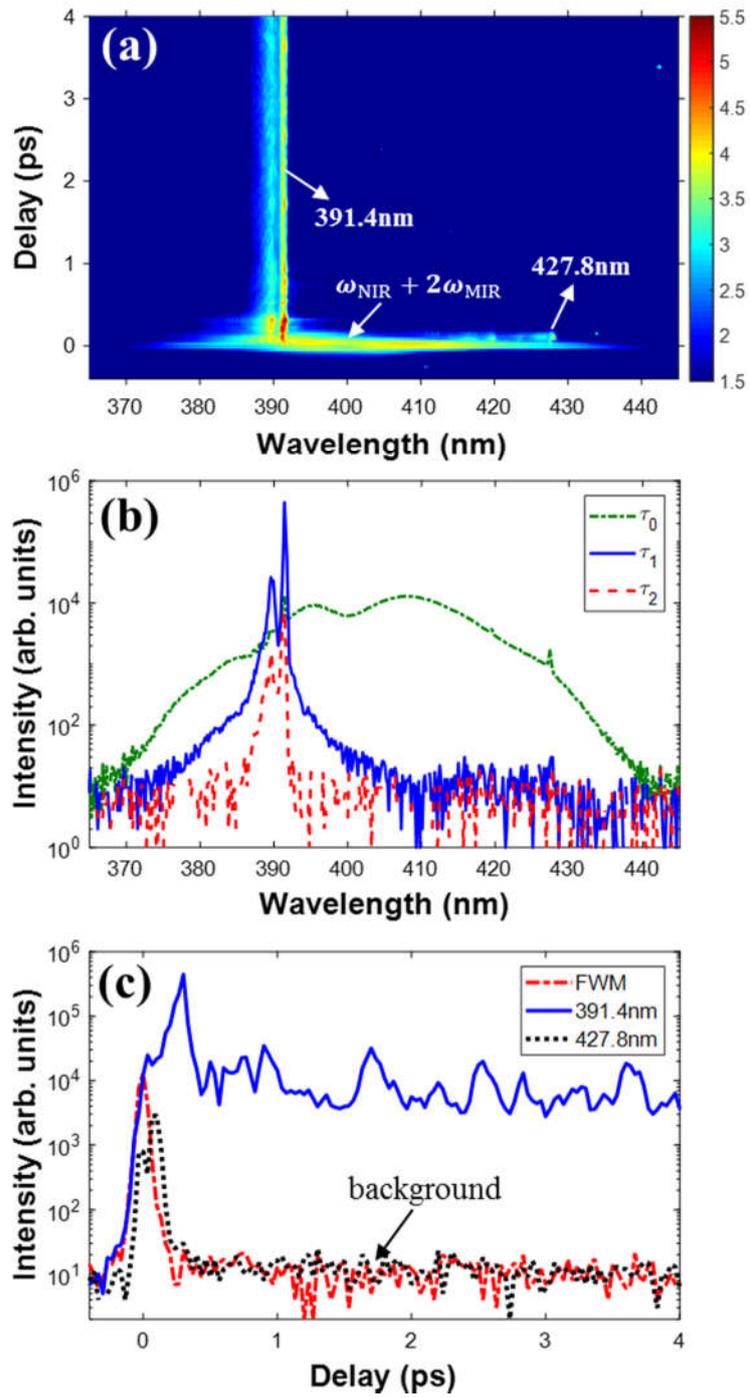

Fig. 3

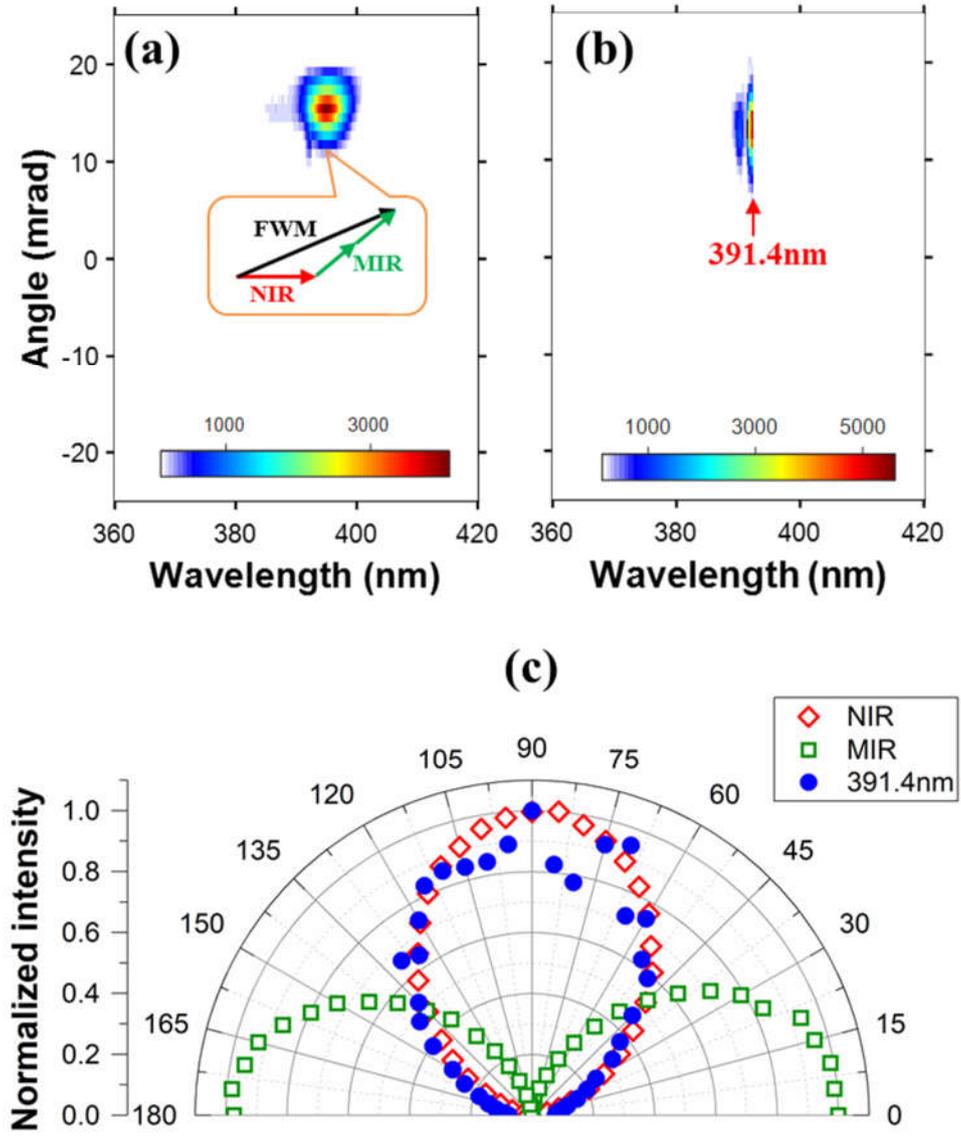

Fig. 4

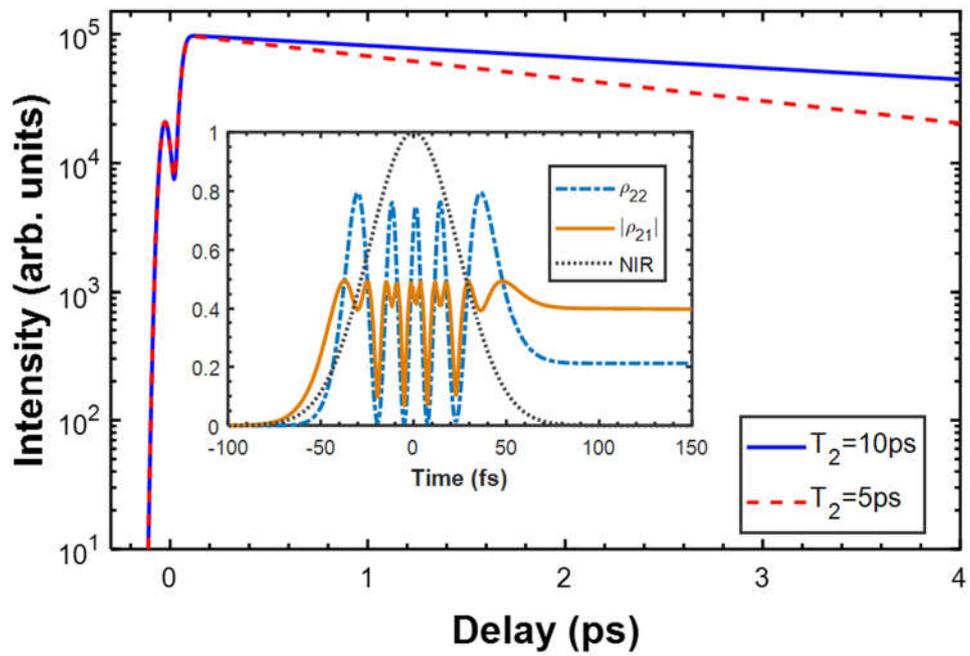

Fig. 5

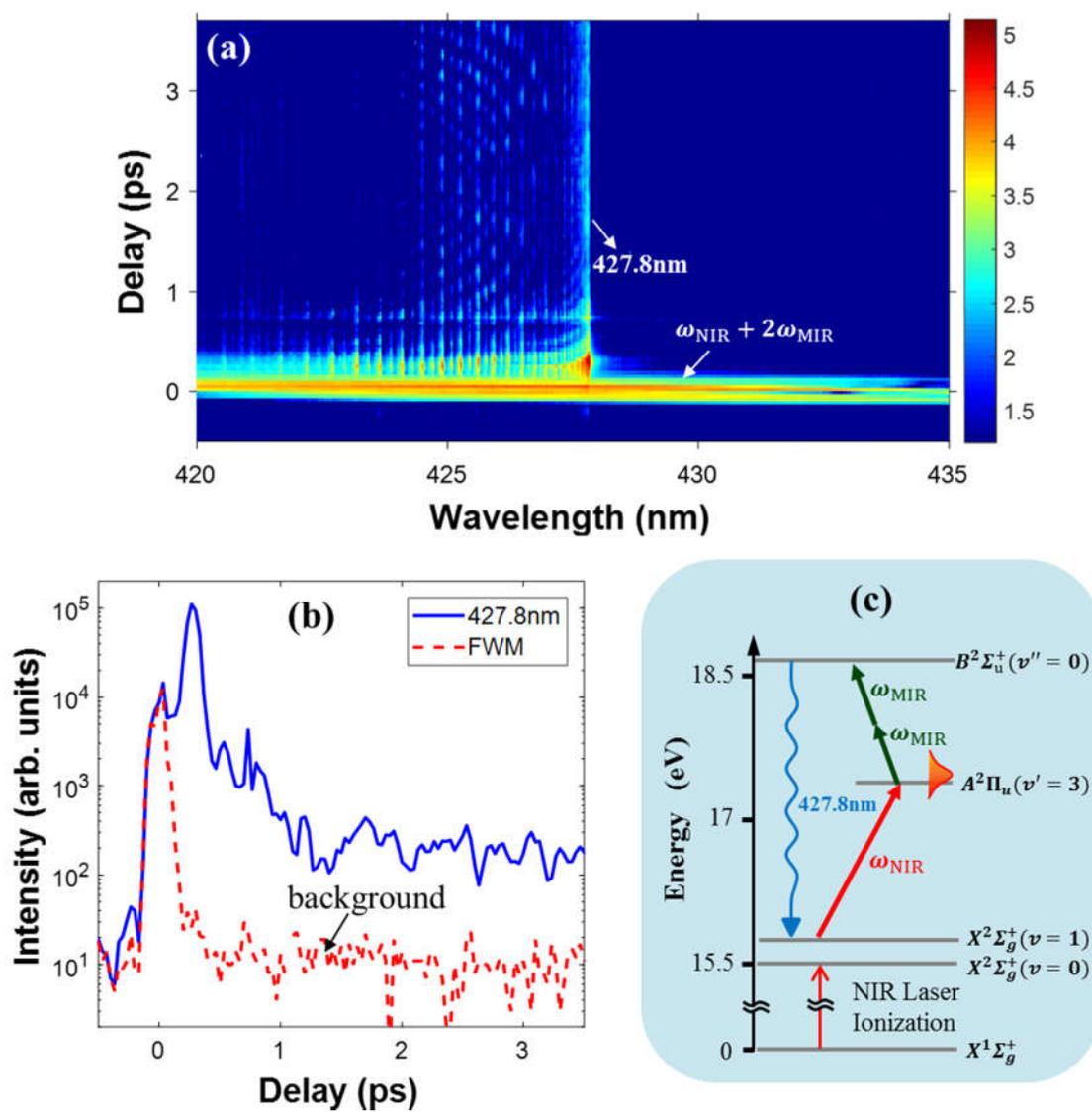